\documentclass[useAMS,usenatbib]{mn2e}
\usepackage{epsfig}
\usepackage{graphics}
\usepackage{longtable,mnlt}
\usepackage{lscape}

\def\SUNSs{SCUBA-2 Unbiased Nearby Stars survey}
\def\DEBRIS{Disc Emission via a Bias-free Reconnaissance in the Infrared/Sub-millimetre}

\title[Target selection for the SUNS/DEBRIS surveys]{Target selection for the SUNS and DEBRIS surveys for debris discs in the solar neighbourhood}
\author[N. M. Phillips et al.]{\begin{minipage}{\textwidth}N. M. Phillips$^1$, J. S. Greaves$^2$, W. R. F. Dent$^3$, B. C. Matthews$^4$ W. S. Holland$^3$,\\ M. C. Wyatt$^5$, B. Sibthorpe$^3$\end{minipage}\\
\\
$^1$Institute for Astronomy (IfA), Royal Observatory Edinburgh, Blackford Hill, Edinburgh, EH9 3HJ\\
$^2$School of Physics and Astronomy, University of St. Andrews, North Haugh, St. Andrews, Fife, KY16 9SS\\
$^3$UK Astronomy Technology Centre (UKATC), Royal Observatory Edinburgh, Blackford Hill, Edinburgh, EH9 3HJ\\
$^4$Herzberg Institute of Astrophysics (HIA), National Research Council of Canada, Victoria, BC, Canada\\
$^5$Institute of Astronomy (IoA), University of Cambridge, Madingley Road, Cambridge, CB3 0HA}
\begin{document}

\date{Accepted 2009 September 2. Received 2009 July 27; in original form 2009 March 31}

\pagerange{\pageref{firstpage}--\pageref{lastpage}} \pubyear{2009}

\maketitle

\label{firstpage}

\begin{abstract}
Debris discs -- analogous to the Asteroid and Kuiper-Edgeworth belts in the Solar system -- have so far mostly been identified and studied in thermal emission shortward of 100$\,\mu{\rm m}$. The {\it Herschel} space observatory and the SCUBA-2 camera on the James Clerk Maxwell Telescope will allow efficient photometric surveying at 70 to 850$\,\mu{\rm m}$, which allow for the detection of cooler discs not yet discovered, and the measurement of disc masses and temperatures when combined with shorter wavelength photometry. The SCUBA-2 Unbiased Nearby Stars (SUNS) survey and the DEBRIS {\it Herschel} Open Time Key Project are complimentary legacy surveys observing samples of $\sim\!500$ nearby stellar systems. To maximise the legacy value of these surveys, great care has gone into the target selection process. This paper describes the target selection process and presents the target lists of these two surveys.
\end{abstract}


\section{Introduction}

The solar neighbourhood is an ideal testing ground for the study of debris
discs and planetary systems. Proximity maximises dust mass sensitivity
and can allow systems to be spatially resolved. Systems near the
Sun span a wide range of stellar parameters e.g., mass, age, metallicity,
multiplicity. Whilst determining these parameters may not be easy, the
diversity included in volume limited samples makes them ideal for legacy
surveys where one may wish to investigate trends as a function of many system
parameters.

This paper presents five all-sky volume limited samples of nearby
stellar systems with main-sequence primaries of spectral type A,F,G,K,M.
These form the basis of the target lists of two complimentary surveys for
debris discs using the SCUBA-2 \citep[Submillimetre Common User Bolometer
Array 2;][]{hol03,aud04} camera on the James Clerk Maxwell Telescope (JCMT), and the
{\it Herschel} space observatory \citep[][]{pil08}.

The \SUNSs\ \citep[SUNS;][]{mat07} is a large flux-limited survey of 500
systems at $850\,\mu{\rm m}$. The target flux RMS is $0.7\,{\rm mJy}/{\rm beam}$, equal to the
extragalactic confusion limit of the JCMT at $850\,\mu{\rm m}$. Shallow
$450\,\mu{\rm m}$ images of varying depth will be obtained simultaneously, and
deep images at $450\,\mu{\rm m}$ will be proposed to follow-up
$850\,\mu{\rm m}$ detections.

The \DEBRIS\ (DEBRIS) {\it Herschel} Open Time Key Program
will image 446 systems (356 in common with SUNS) at 110 and $170\,\mu{\rm m}$ using the PACS
\citep[Photodetector Array Camera and Spectrometer;][]{pog08} instrument,
with follow-up of around 100 systems at 250, 350 and $500\,\mu{\rm m}$ using the SPIRE
\citep[Spectral and Photometric Imaging Receiver;][]{gri08} instrument. This survey is
primarily driven by the $110\,\mu{\rm m}$ band, which has the highest dust mass
sensitivity for cold discs such as the Kuiper-Edgeworth belt of our Solar System.
The intended flux RMS at $110\,\mu{\rm m}$ is $1.2\,{\rm mJy}/{\rm beam}$, which is twice the predicted
extragalactic confusion limit. $170\,\mu{\rm m}$ images are taken
simultaneously with a predicted RMS of $1.7\,{\rm mJy}/{\rm beam}$, equal to the predicted extragalactic
confusion limit in this band.

The primary goals of these surveys are statistical: In general how do debris
disc properties vary with stellar mass, age, metallicity, system
morphology (multiplicity, component masses, separations), presence of planets
etc. To be able to answer so many questions, and to minimise the risk of
unforeseen selection effects, large samples and simple, clearly defined target
selection criteria are required. Volume limited samples satisfy these
requirements, and as well as maximising the proximity of the targets, the
stars nearest the Sun are very widely studied. For example, nearby stars are
the main targets of radial velocity, astrometry and direct imaging planet
searches. The majority of SUNS and DEBRIS targets also have photometry at 24 and
$70\,\mu{\rm m}$ from the MIPS (Multiband Imaging Photometer and Spectrometer)
instrument on the {\it Spitzer} space telescope, which ceased operation at the
end of March 2009. This large spectral coverage, from 24 to
$850\,\mu{\rm m}$ for over 300 systems will be an incredible resource for
 detailed spectral energy distribution modelling of systems with debris discs.

Given that we are considering the closest systems to the Sun, substantial
effort was required to compile the samples presented here. Late M-type
stars within $10\,{\rm pc}$ are still being discovered \citep[e.g.][]{hen06},
and complete homogeneous datasets covering the spectral type and distance
ranges we consider do not exist. We have tried to make our sample selection
using the most complete and accurate data available at the time of the DEBRIS
proposal submission in October 2007.

\section{Selection Criteria}
Our systems all have primaries (defined here as the component with the
brightest visible magnitude) which we believe are main-sequence (i.e. hydrogen
burning) stars. The sample is split into 5 volume limited subsamples based on
spectral type: A, F, G, K, M. In the rest of this paper we use the term
``X-type system'' to mean ``system with X-type primary''. Using separate
subsamples is necessary due to the steep nature of the stellar mass function,
which for example means that a single volume limited sample would contain over
100 times as many M-type systems as A-type systems. The choice of using
spectral types to split the sample, rather than stellar mass, is purely
practical as, with the exception of certain binary systems, stellar masses
cannot be directly determined observationally. Using spectral types does however have
the effect that the subsamples cover quite different ranges in
logarithmic mass space.

The early type, upper mass, limit of A0 is chosen as stars of earlier type are too
rare in the solar neighbourhood to build a suitably large sample. A
conservative late type limit of M7.0 was chosen to avoid the inclusion of any
brown dwarfs, and also to improve the completeness of the M-type sample.
M-type stars span the largest $\log M$ range of any of our spectral classes,
so making a cut at M7.0 will not restrict the statistical usefulness of the
sample.

We do not discriminate against multiple star systems, and they are included naturally within the volume limits. We consider common proper motion stars (with compatible parallax where available) as members of the same system, with no specific limit on the binary separation. We have not gone so far as to consider stars with common space motion but large ($\gg\!\! 1^\circ$) angular separation as systems. This definition of system membership was primarily chosen for convenience of target selection, but fits well with the statistical goals of these surveys. With the exception of stars in moving groups, each system can be considered to represent a different point in age and composition. The fact that several interesting objects (e.g. with known IR excess, or planets) are considered here as secondaries does not affect the statistical usefulness of the sample, although it has the disadvantage that such objects may not be observed by these surveys (see below).

The number of systems in each subsample was determined by the selection
criteria for SUNS, which required 100 systems in each subsample in the
declination range $-40^\circ < \delta < +80^\circ$. Hence the all-sky samples
presented here contain roughly 123 ($100 \times 2 / (\sin80^\circ + \sin40^\circ)$) systems
each. The SUNS sample sizes were chosen to allow detection rates for various
subsets e.g. planet hosts to be distinguished \citep[see][]{mat07}.

% figure position dictated by proofs
\begin{figure}
\centerline{\includegraphics{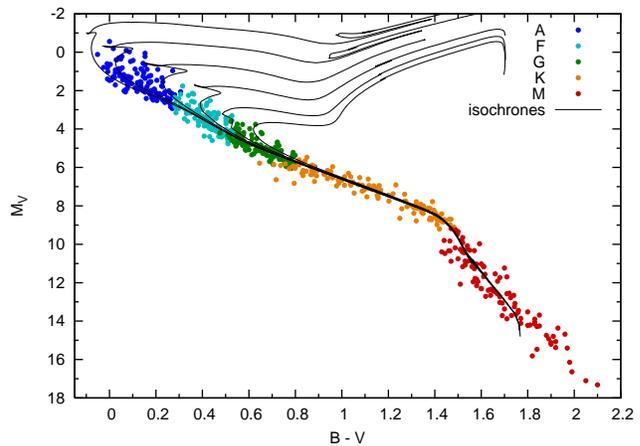}}
\caption{
	Johnson B,V absolute colour-magnitude diagram for system primaries.
	Overlaid are $[{\rm Fe}/{\rm H}]=0.0$, $[\alpha/{\rm Fe}]=0.0$
	isochrones from the Dartmouth
	Stellar Evolution Database \citep{dot08} with ages of $0.25$,
	$0.5$, 1, 2, 4, $8\,{\rm Gyr}$ (with turn-offs going from left to
	right). The photometry is mostly converted from Tycho photometry
	(Tycho-2 or TDSC) using transformations for unreddened main-sequence
	stars. For most M-type targets Johnson B,V photometry from various
	sources was used (see text). Note that primaries in some close
	binaries are not individually resolved in this photometry.
}
\label{fig:colmag_isochrones}
\end{figure}

\begin{figure}
\centerline{\includegraphics{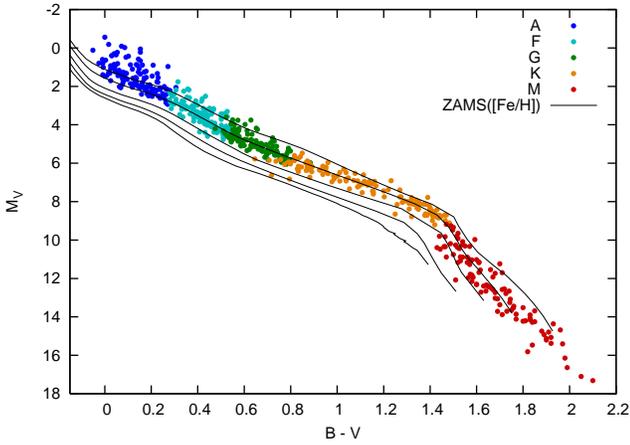}}
\caption{
	Johnson B,V absolute colour-magnitude diagram for system primaries
	as in figure 1. Overlaid with zero age main-sequences (ZAMS) for stars
	from $0.2\,M_\odot$ upwards with
	$[{\rm Fe}/{\rm H}]=+0.5,0.0,-0.5,-1.0,-2.0$
	(from top to bottom). The
	ZAMS curves are produced from $[\alpha/{\rm Fe}]=0.0$, $Y=0.245+1.6Z$
	evolutionary tracks from the Dartmouth
	Stellar Evolution Database \citep{dot08}, with values taken at
	2\% of the total lifetime of the stars.
}
\label{fig:colmag_zams}
\end{figure}

\begin{table*}
\caption{
	Summary of subsample properties. $d_{\rm max}$ and $N_{\rm tot}$ are the maximum distance and number of stars in each subsample. $\rho$ is the volume number density of systems, $\rho=N_{\rm tot}/d_{\rm max}^3 \pm \rho/\sqrt{N_{\rm tot}}$. ${\rm Med}(T_{\rm eff})$ is the median $T_{\rm eff}$, and $\sigma_{T_{\rm eff}}$ is the standard deviation of $T_{\rm eff}$ within each subsample. $N_{\rm planet}$ is the number of systems where one or more stars are listed as planet hosts in the {\tt exoplanet.eu} database (27 July 2009). $N_{\rm debris}$ is the number of systems containing a currently detected debris disc (or other indistinguishable IR excess) as indicated by any of \citet{rhe07,bei06b,su06,tri08}. $N_{\rm SUNS}$ and $N_{\rm DEBRIS}$ are the numbers of systems from this paper included in the SUNS and DEBRIS surveys respectively.
}
\begin{tabular}{c|c|c|c|c|c|c|c|c|c}
\hline
Subsample & $d_{\rm max}$ & $N_{\rm tot}$ & $\rho$ & ${\rm Med}(T_{\rm eff})$ & $\sigma_{T_{\rm eff}}$ & $N_{\rm planet}$ & $N_{\rm debris}$ & $N_{\rm SUNS}$ & $N_{\rm DEBRIS}$\\
 & (pc) & & (${\rm pc}^{-3}$) & (K) & (K) & & & & \\
\hline
A & 45.5 & 130 & $0.0014 \pm 0.0001$ & 8133 & 748 &  2 & 24 & 100 & 83 \\
F & 24.1 & 130 & $0.0093 \pm 0.0008$ & 6360 & 343 &  6 & 21 & 100 & 94 \\
G & 21.3 & 125 & $0.0129 \pm 0.0012$ & 5628 & 249 & 13 & 10 & 100 & 89 \\
K & 15.6 & 127 & $0.0335 \pm 0.0030$ & 4461 & 499 &  5 &  5 & 100 & 91 \\
M & 8.58 & 117 & $0.1855 \pm 0.0171$ & 3175 & 288 &  5 &  1 & 100 & 89 \\
% sigma_teff / Med(T_eff) = 0.092, 0.054, 0.044, 0.112, 0.091 (A-M)
\hline
Total & & 629 & & & & 31 & 61 & 500 & 446
\end{tabular}
\end{table*}

The DEBRIS target list comprises the nearest systems presented here (all-sky),
subject to a cut in the predicted $110\,\mu{\rm m}$ cirrus confusion level
towards each system. The confusion prediction was taken from the Herschel Confusion Noise Estimator (HCNE), which is part of the Herschel Observation Planning Tool (HSPOT). Systems with total predicted confusion for point-source
detections greater than $1.2\,{\rm mJy}/{\rm beam}$, corresponding to twice the predicted extragalactic confusion limit, were rejected. To maximise the number of systems observed, DEBRIS will not image secondary components in systems where they will not fit in the PACS point-source field of view (FoV) ($150''\times50''$ with unconstrained orientation) with the primary. This will affect between 20 and 49 systems depending on the actual field orientations.

The SUNS target list is simply the nearest 100 systems in each subsample here
which have $-40^\circ < \delta < +80^\circ$ (with this sample it does not make any
difference whether the cut is made in B1950 or J2000/ICRS equinox declination,
but J2000/ICRS should be assumed). The large ($\sim600''\times600''$) FoV of
SCUBA-2 means that a maximum of 13 systems will have components not observed with
the primary star.

Initially, it had been proposed to only include systems with primaries of
spectroscopic luminosity classes V and IV-V. This criterion was retained for G,
K and M classes, but was relaxed for A and F type stars, where there is not a
simple relationship between luminosity class and evolutionary stage
\citep[e.g.,][]{gray01a,gray01b}. Candidates for the A and F samples (and other
candidates without accurately known luminosity classes) were evaluated using
their position on a Johnson B,V absolute colour-magnitude diagram. Figs
\ref{fig:colmag_isochrones} and \ref{fig:colmag_zams}
show such diagrams for the final sample overlaid with solar composition
isochrones and zero age main sequences (ZAMS) for metallicities from
$+0.5$ to $-2.0$. A certain amount of leeway had to be allowed for unknown
metallicity, and uncertainties in photometry (e.g. unresolved secondaries in
close binaries) and parallax.

\section{Sources of data}

\subsection{Parallaxes}

{\it Hipparcos}-based parallaxes were taken from {\it Hipparcos, the New Reduction of the Raw Data} \citep[HIPnr;][]{HIPnr} and several papers which applied special analysis to multiple systems \citep[the {\it General Notes} issued with the original {\it Hipparcos} catalog (HIPgn, Perryman et al. 1997);][]{fal99,sod99,fab00}. Parallaxes from HIPnr were used unless one of the other resources had a lower uncertainty. In cases where more than one of the other resources provided a parallax for the same {\it Hipparcos} system, we have have taken the parallax from the first resource in the order: \citet{fab00,sod99,fal99}; HIPgn. {\it Hipparcos} parallaxes from multiple resources for the same {\it Hipparcos} system were not averaged in any way to avoid underestimating the uncertainty in the averaged values, as they have all been reduced from the same data.

The other large parallax resource used was the 4th edition of the Yale General Catalog of Trigonometric Parallaxes \citep[GCTP or YPC;][]{YPC}, which contains approximately 2300 systems not measured by {\it Hipparcos} due to the magnitude limit of $V\sim12$ and the targeted nature of the {\it Hipparcos} astrometry mission.

In addition, for many M dwarfs, parallaxes from several smaller papers were used \citep[e.g.,][]{hen06,jao05,cos05,her98,ben99,wei99,duc98}, as well as some unpublished values from the RECONS consortium (Henry, private communication).

Where reliable parallaxes from multiple independent sources, or separate parallaxes for individual components in a system, are available, we take an uncertainty weighted average:
$$\pi_{\rm adopted} = {\sum_i \pi_i/\sigma_i^2\over\sum_i 1/\sigma_i^2}
\quad{\rm and}\quad
\sigma_{\rm adopted} = \sqrt{1\over\sum_i 1/\sigma_i^2}$$

Two or more parallaxes were used for 81\% of systems, and three or more were used for 7\% of systems. These cases are mostly due to overlap with {\it Hipparcos}- and ground-based (e.g. YPC) parallaxes.

\subsection{Spectral Types}

For A-K type stars we have used spectral types from \citet{gray03,gray06} where they were available. Gray et al. have been obtaining spectra and determining spectral types and stellar parameters ($T_{\rm eff}$, $[M/H]$, $\log g$) for stars considered to be within $25\,{\rm pc}$ and of spectral type earlier than M0 or with no spectral type in the {\it Hipparcos} catalogue \citep[HIP;][]{HIP}. For stars without published Gray et al. types we have used types from the Michigan Catalogue of HD stars \citep{houk1,houk2,houk3,houk4,houk5}, which includes all HD stars south of $\delta_{\rm B1900} = +05^\circ$. If types from neither Gray et al. or Houk et al. were available, we have fallen back on types in compilations such as the 5th revised edition of the Bright Star Catalogue \citep[BSC5;][]{BSC5}, HIP, or the 2nd edition of the Catalog of Components of Double \& Multiple stars \citep[CCDM;][]{CCDM}. These fall-back types are not considered to be accurate, and were largely ignored in the selection process in favour of photometry.

% moved here to reflect proof layout
\begin{table*}
\caption{
	Reference abbreviations used in the text and tables. CDS is Centre de Donn\'ees astronomiques de Strasbourg. For HIPnr we have used the data on the CDROM published with the book, as it had not been added to the CDS at the time.
}
\begin{tabular}{c|l|l}
\hline
Abbreviation & CDS catalogue(s) & Reference \\
\hline
2MASS  & II/246                     & 2MASS Point Source Catalogue \citep{2MASS} \\
BSC5   & V/50                       & Bright Star Catalogue, 5th Revised Ed. \citep{BSC5} \\
CCDM   & I/274                      & Catalogue of Components of Double \& Multiple stars \citep{CCDM} \\
CNS3   & V/70A                      & Catalogue of Nearby Stars, Preliminary 3rd Version \citep{CNS3} \\
HIP    & I/239                      & {\it Hipparcos} Main Catalogue \citep{HIP} \\
HIPgn  & I/239                      & {\it Hipparcos} General Notes \citep{HIP} \\
HIPnr  & I/311*                     & {\it Hipparcos}, the New Reduction of the Raw Data \citep{HIPnr} \\
LHS    & I/279                      & Revised Luyten Half-Second catalogue \citep{bak02} \\
NLTT   & J/ApJ/582/1011             & Revised NLTT Catalog \citep{sal03} \\
PPM    & I/\{146, 193, 206, 208\}   & Positions and Proper Motions catalogue \citep{PPM1,PPM2,PPM3} \\
RECXX  &                            & RECONS unpublished parallaxes (Henry, private communication) \\
SCR    & J/AJ/\{129/413, 130/1658,  & SuperCOSMOS-RECONS \citep{sub05a,sub05b,fin07} \\
       & 133/2898\}                 & \\
TDSC   & I/276                      & Tycho Double Star Catalogue \citep{TDSC} \\
TRC    & I/250                      & Tycho Reference Catalogue \citep{TRC} \\
TYC    & I/239                      & Tycho catalogue \citep{TYC} \\
TYC2   & I/259                      & Tycho-2 catalogue \citep{TYC2} \\
YPC    & I/238A                     & Yale Parallax Catalogue, 4th ed. \citep{YPC} \\
WDS    & B/wds                      & Washington Visual Double Star Catalog \citep{WDS} \\
ben99  &                            & \citet{ben99} \\
bes90  &                            & \citet{bes90} \\
cos05  &                            & \citet{cos05} \\
dea05  & J/A+A/435/363              & Southern Infrared Proper Motion Survey \citep[SIPS, ][]{dea05} \\
duc98  &                            & \citet{duc98} \\
egg74  &                            & \citet{egg74} \\
egg79  &                            & \citet{egg79} \\
egg80  &                            & \citet{egg80} \\
fab00  & J/A+AS/144/45              & \citet{fab00} \\
fal99  & J/A+AS/135/231             & \citet{fal99} \\
gou04  & J/ApJS/150/455             & \citet{gou04} \\
gray03 & J/AJ/126/2048              & \citet{gray03} \\
gray06 & J/AJ/132/161               & \citet{gray06} \\
jao05  &                            & \citet{jao05} \\
hen06  &                            & \citet{hen06} \\
haw95  & III/198                        & Palomar/MSU survey (North) \citep{haw95} \\
haw96  & III/198                        & Palomar/MSU survey (South) \citep{haw96} \\
houk   & III/\{31B, 51B, 80, 133, 214\} & Michigan Catalogue of HD stars \citep{houk1,houk2,houk3,houk4,houk5} \\
her98  &                                & \citet{her98} \\
leg92  &                                & \citet{leg92} \\
rod74  &                                & \citet{rod74} \\
sod99  & J/A+A/341/121                  & \citet{sod99} \\
wei91  &                                & \citet{wei91} \\
wei96  &                                & \citet{wei96} \\
wei99  &                                & \citet{wei99} \\
\hline
\end{tabular}
\end{table*}

For $\sim$K5 and later stars we have generally used spectral types from the Palomar/MSU Nearby-Star Spectroscopic Survey \citep[PMSU;][]{haw95,haw96}, which provides spectral types for almost all late-type stars in the 3rd Catalogue of Nearby stars \citep[CNS3;][]{CNS3}. A large number of nearby M dwarfs also have measured spectral types in the system or \citet{kir91}; however we have chosen to use PMSU types wherever possible for homogeneity. The difference between PMSU and Kirkpatrick et al. types is rarely more than one subtype. For newly discovered nearby M dwarfs not included in the PMSU, types in the Kirkpatrick et al. system \citep[e.g. from][]{hen06} have been adopted.

\subsection{Photometry}

Whilst distance and spectral type are our primary selection parameters, it was necessary to use photometry both for determining luminosity and when determining spectral class where only low accuracy spectral types were available. As distinguishing between dwarfs and giants for K/M type stars is very simple and because we had accurate spectral types for almost all candidates later than K5 (see above), photometry was only needed for the selection of systems on the G/K boundary and earlier. All of these candidates are bright enough to have sufficiently accurate photometry in the Tycho-2 catalogue \citep{TYC2}, the Tycho Double Star Catalogue \citep[TDSC;][]{TDSC}, or the Tycho catalogue \citep{TYC}. Where there has been a need to convert between Tycho and Johnson photometry we have used the relationships in \citet{TYC2}.

\begin{figure}
\centerline{\includegraphics{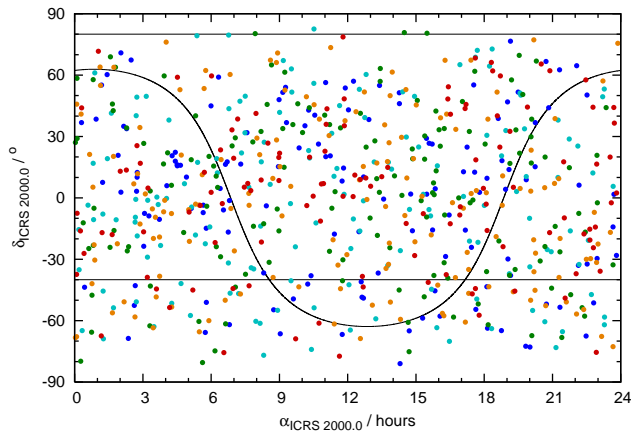}}
\caption{
	Distribution of all systems in ICRS equatorial coordinates. The SUNS declination limits of $+80^\circ$ and $-40^\circ$, and the Galactic plane are shown.
}
\label{fig:positions_all}
\end{figure}

\begin{figure}
\centerline{\includegraphics{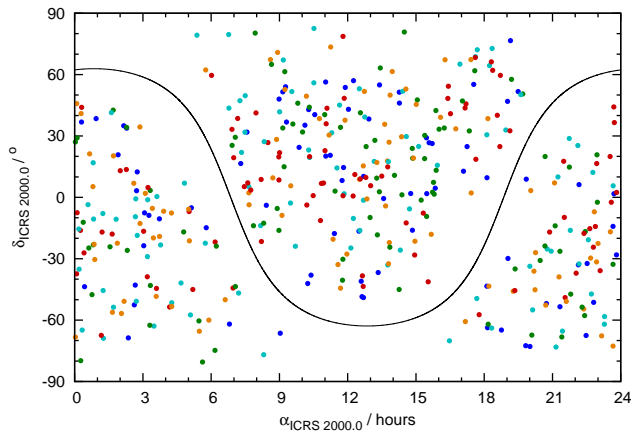}}
\caption{
	Distribution in ICRS equatorial coordinates of the 446 systems in the DEBRIS survey. The cut in predicted cirrus confusion means that there are few systems near the Galactic plane.
}
\label{fig:positions_debris}
\end{figure}

\begin{figure}
\centerline{\includegraphics{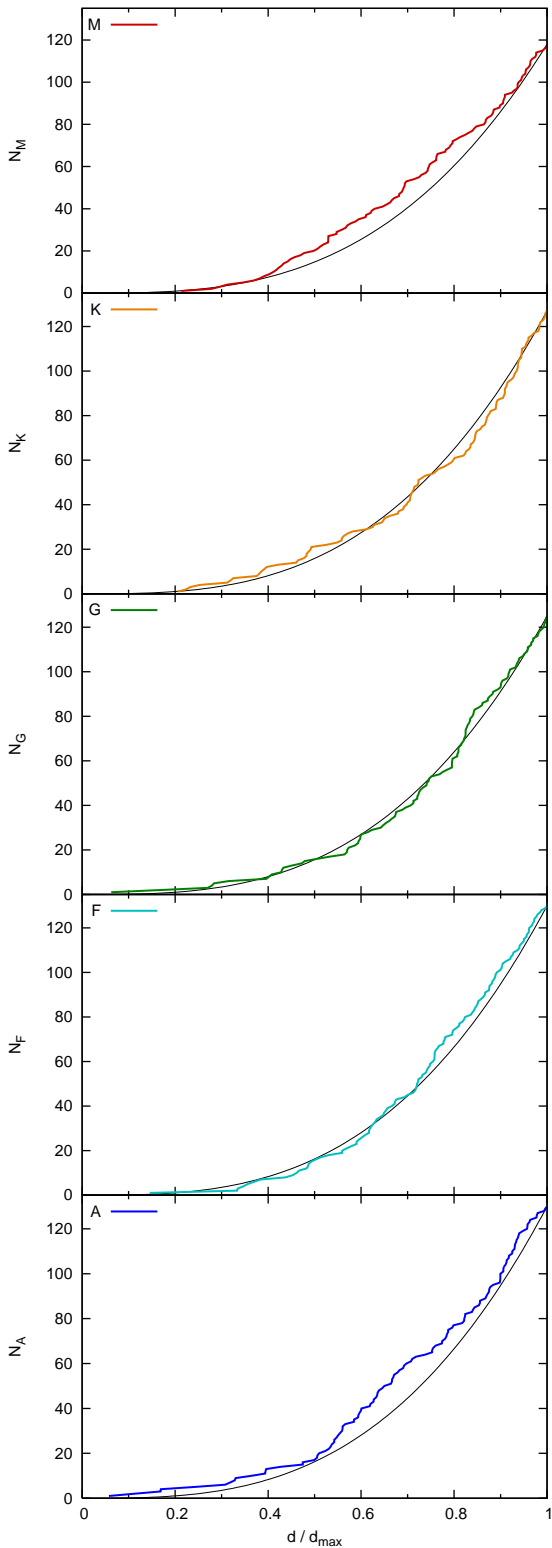}}
\caption{
	Number of included systems in each subsample as a function of distance ($d_{\rm max}=8.58,15.6,21.3,24.1,45.5\,{\rm pc}$ for M,K,G,F,A). For comparison the line $N=N(d_{\rm max})\left(d\over d_{\rm max}\right)^3$ is shown. Note that the F,G,K subsamples fit well indicating no completeness trend with distance. The M subsample is likely incomplete beyond $\sim\!6\,{\rm pc}$.
}
\label{fig:completeness}
\end{figure}

\begin{figure}
\centerline{\includegraphics{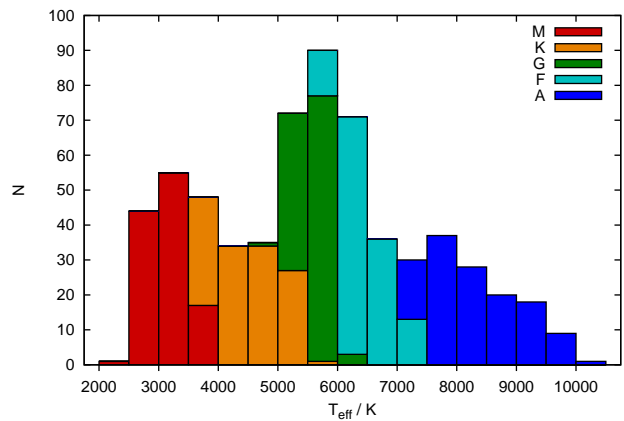}}
\caption{
	Histogram of number of primaries in $500\,{\rm K}$ $T_{\rm eff}$ bins. Contributions from each spectral type subsample are shown in colour. For A-K stars $T_{\rm eff}$ was derived from $(B_{\rm T}-V_{\rm T})$ (or $(B_{\rm J}-V_{\rm J}$ in a few cases where Tycho photometry was not available) using a polynomial fit against $T_{\rm eff}$ values from \citet{gray03,gray06} (see Fig. \ref{fig:teff_bv}). $(B_{\rm T}-V_{\rm T})$ was used in preference to the more accurate temperature indicator $(V-K_{\rm s})$, as components are resolved at very small separations in Tycho-2/TDSC photometry. For M-type stars $T_{\rm eff}$ was derived from our adopted spectral type using $T_{\rm eff}$ values from \citet{rei05}.
}
\label{fig:teff_hist}
\end{figure}

\begin{figure}
\centerline{\includegraphics{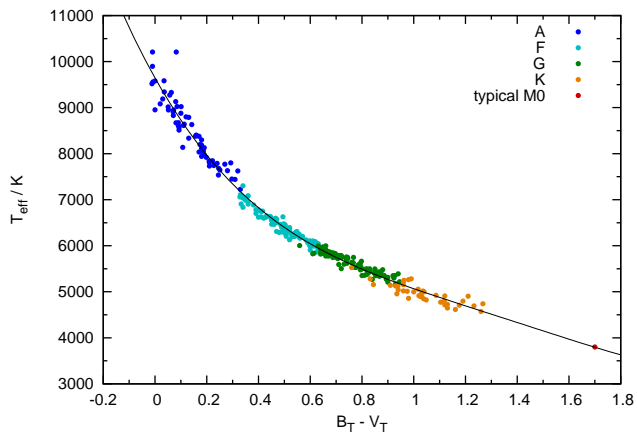}}
\caption{
	\citet{gray03,gray06} $T_{\rm eff}$ vs. $(B_{\rm T}-V_{\rm T})$ for primary stars in our sample, with 4th order polynomial fit. This fit was used to generate $T_{\rm eff}$ values for all A-K primaries for Fig. \ref{fig:teff_hist}. A point for a typical M0 type star at $(1.70,3800)$ was added to the fit to make it tie in with $T_{\rm eff}$ values for M type stars derived from spectral types using relationships in \citet{rei05}.
}
\label{fig:teff_bv}
\end{figure}

\subsection{Astrometry}

Accurate positions and proper motions were necessary both for matching entries in the various catalogues used, and for finding common proper motion companions. Where possible, astrometry from \citet{sal03,gou04,dea05,sub05a,sub05b,fin07,hen06,jao05} have been used. For stars not included or not resolved in these, we have used astrometry from the TDSC; Tycho-2; the Tycho Reference Catalogue \citep[TRC;][]{TRC}; Tycho; \citet{bak02}; the Positions and Proper Motions catalogue \citep[PPM;][]{PPM1,PPM2,PPM3}; or the CCDM (in order of decreasing preference).

\section{Components of Multiple Systems}

We have undertaken several steps to maximise the accuracy of the selection of components in multiple systems.

Using the database we have constructed for the purposes of the target selection, we have searched for stars with common proper motion to candidate targets. This not only yielded secondary stars which we had not previously identified, but also showed some candidates to be secondaries of other stars. In cases where common proper motion companions have independent parallax measurements, these have been checked to be compatible. Other common proper motion companions have been identified from literature, although a systematic literature search for such companions has not been performed.

We have performed a complete check of all components listed in the CCDM as being in the CCDM systems of our targets. In many cases components listed in the CCDM are not physically associated (e.g. do not have common proper motion) with the target system. Many CCDM components have cross-identifications with other catalogues, so determining whether they have common proper motion is straightforward. For those without cross-identifications, or without accurate astrometry in other catalogues, only the astrometry in the CCDM could be used.

The process for determining system membership of CCDM components consisted of an automated search for components using the 2MASS Point Source Catalogue \citep{2MASS}, and the Tycho/Tycho-2 catalogues, followed by manual inspection of 2MASS and Schmidt survey images, as well as comparison with the Washington Double Star catalogue \citep{WDS} in many cases. CCDM components found not to be comoving with the target systems, or not identified at all, are not included in the sample presented here, but are listed in comments in table~8.

\section{Sample Properties}

Overall properties of the subsamples are presented in table~1, including the numbers of systems containing stars with detected planets and debris discs. Figs \ref{fig:positions_all} and \ref{fig:positions_debris} show the distribution of systems on the sky.

\subsection{Completeness}
In Fig. \ref{fig:completeness}, we show the number of systems as a function of distance for each of our subsamples. The F,G and K subsamples very closely follow a cubic law, indicating that we are justified to assume they are isotropically and homogeneously distributed in the relevant volumes and that we have no selection effects as a function of distance. For the M subsample there is almost certainly incompleteness at distances beyond $\sim\!6\,{\rm pc}$ \citep[see e.g.][]{hen94} which will mostly affect the latest type stars. The deviation of the A subsample from the cubic law is likely a combination of a slight lack of systems towards the Galactic poles at the largest distances, and correlation between system positions due to the young age of A stars.

\subsection{Temperature Distribution}
As our sample was split into subsamples based on spectral class, we expected to have a good coverage of effective temperature of primary stars from about 2500 to $10000\,{\rm K}$ (M7-A0 types). Fig. \ref{fig:teff_hist} shows the distribution of $T_{\rm eff}$ for primary stars in our sample in $500\,{\rm K}$ bins. The colours in the plot indicate the contributions from the five A-M subsamples. For A-K stars, $T_{\rm eff}$ was computed from $(B_{\rm T}-V_{\rm T})$ using a fit to $T_{\rm eff}$ for stars in our sample from \citet{gray03,gray06}. $(B_{\rm T}-V_{\rm T})$ was chosen as opposed to other photometric colours such as $(B_{\rm J}-V_{\rm J})$ or $(V_{\rm T}-K_{\rm s})$, as accurate homogeneous $B_{\rm T}$ and $V_{\rm T}$ photometry that is resolved down to separations of $<0.5''$ is available for almost all of our A-K primaries from the Tycho-2 and Tycho Double Star (TDSC) catalogues. The fit of $(B_{\rm T}-V_{\rm T})$ to Gray et al.'s $T_{\rm eff}$ values is show in Fig. \ref{fig:teff_bv}. A fourth order least-squares polynomial fit was obtained:

\begin{eqnarray}
T_{\rm eff}/{\rm K} &=& (9646.15 \pm 37.6)\nonumber\\
&&- (10018.4 \pm 354.4)(B_{\rm T}-V_{\rm T})\nonumber\\
&&+ (9056.19 \pm 963.2)(B_{\rm T}-V_{\rm T})^2\nonumber\\
&&- (4424.10 \pm 950.5)(B_{\rm T}-V_{\rm T})^3\nonumber\\
&&+ (807.378 \pm 302.8)(B_{\rm T}-V_{\rm T})^4\nonumber
\end{eqnarray}

This agrees well with the fit of \citet{ram05} with $[{\rm Fe}/{\rm H}]=0.0$ for their range of validity of $0.344 < (B_{\rm T}-V_{\rm T}) < 1.715$. Our RMS of residuals is $150.7\,{\rm K}$ for 302 stars, which is higher than that of \citet{ram05} ($104\,{\rm K}$ for 378 stars), as we cover a larger temperature range, have not used $[{\rm Fe}/{\rm H}]$ as a fit parameter, and have not accounted for interstellar reddening (although this should be almost negligible for our nearby star sample).

For M-type stars we determined $T_{\rm eff}$ simply from our adopted spectral type using values from \citet{rei05}. The above photometric fit for A-K stars included a point representative of a typical M0 type star at $(B_{\rm T}-V_{\rm T})=1.70, T_{\rm eff}=3800\,{\rm K}$ to make the fit consistent with our M star temperatures at the K/M boundary.

The peak in the $T_{\rm eff}$ distribution at about $5700\,{\rm K}$ is due to the G and F spectral types covering a narrow range in $T_{\rm eff}$. Indeed, in retrospect, there would be justification for treating F and G types as a single spectral type sample.

\section{Catalogue}
Table 2 lists the reference abbreviations used throughout this paper and in the other tables.
Tables~3-6 define the sample, and give information used in the selection process. Each system is given an identifier of the form XNNN where X is the spectral class (subsample) and NNN is a zero-padded running number increasing with distance in each subsample. These identifiers are referred to by the acronym UNS, standing for Unbiased Nearby Stars, as in the SUNS survey name.

The choice of name for components is generally in order of preference: HD, HIP, GJ, LHS, NLTT, TYC, PPM, CCDM, other catalogue name, 2MASS. For systems with multiple stars, the first identifier in that order which uniquely identifies the component is used. Where components are not resolved in any catalogues we have used, we just give a single entry and a comment in table~8.

Table~3 lists system properties, including the name of the primary star, our adopted distance, and whether the system is included in the SUNS and DEBRIS surveys.

Table~4 lists the components of systems which are resolved in at least one of the catalogues we have used, and gives positions and proper motions, as well as approximate separation from the primary where this is larger than $1''$. Where two references are listed for a component, the proper motion has been copied from another component in the system, and in several cases the position is computed using a relative position from the CCDM combined with the position of another component.

Tables 5 and 6 list the properties of primary stars in systems, which were used for selection in spectral type and luminosity (spectral type, photometry), and/or in the plots in this paper (photometry, effective temperatures). Table 5 contains the A-K type primaries with Tycho photometry, and effective temperatures from \citet{gray03,gray06} and computed from $(B_{\rm T}-V_{\rm T})$. For the few very bright stars where Tycho photometry is saturated, we give values converted from Johnson $B,V$ photometry. Table 6 contains the M type primaries with spectral types, Johnson $B,V$ photometry, and effective temperatures computed from the spectral type.

Table~7 gives cross identifications for system components in several common catalogues, and table~8 gives comments on various specific systems. Table 8 includes notes for systems where there are unresolved components, or there are components listed in catalogues which we do not consider physically associated with the system.

\section{Acknowledgements}
This research has made use of the SIMBAD and VIZIER databases, operated at CDS, Strasbourg, France. We have made extensive use of the open source MySQL relational database management system. The authors wish to thank Todd Henry for providing unpublished RECONS parallaxes, which helped to refine our selection of M type systems.

%\clearpage

\onecolumn

\small

\begin{longtable}{cr@{ }lr@{}c@{}l@{ $\pm$ }r@{}c@{}lclr@{}c@{}lc@{}c}
\caption{System information: system ID, primary star name, adopted distance and uncertainty ($d=1/\pi \pm \sigma_\pi / \pi^2$), number of parallax measures used, parallax references (see table 2), predicted cirrus confusion noise for point source observed with {\it Herschel}'s PACS instrument at $110\,\mu{\rm m}$, which surveys system is included in (S: SUNS, D: DEBRIS). Note that distance uncertainty is not shown for the 2 systems with unpublished RECONS parallaxes. The distance for UNS G001 ($\alpha$ + Proxima Centauri) does not include any contribution from Proxima, as the parallax difference from the primary is significant.
}\\
\hline
{UNS ID}&\multicolumn{2}{c}{Primary}&\multicolumn{6}{c}{d}&{$N_\pi$}&References&\multicolumn{3}{c}{$C_{{\rm PACS},110}$}&\multicolumn{2}{c}{Surveys}\\
{}&{}&{}&\multicolumn{6}{c}{(pc)}&{}&{}&\multicolumn{3}{c}{(${\rm mJy}/{\rm beam}$)}&{}&{}\\
\hline
\endfirsthead

\caption{System information (continued)}\\
\hline
{UNS ID}&\multicolumn{2}{c}{Primary}&\multicolumn{6}{c}{d}&{$N_\pi$}&References&\multicolumn{3}{c}{$C_{{\rm PACS},110}$}&\multicolumn{2}{c}{Surveys}\\
{}&{}&{}&\multicolumn{6}{c}{(pc)}&{}&{}&\multicolumn{3}{c}{(${\rm mJy}/{\rm beam}$)}&{}&{}\\
\hline
\endhead

\multicolumn{16}{r}{Continued on next page...} \\
\endfoot

\hline
\endlastfoot

M001&HIP&87937&1&.&833&0&.&001&2&YPC,HIPnr,ben99&1&.&29&S&\\
M002&GJ&406&2&.&386&0&.&012&1&YPC&0&.&53&S&D\\
M003&HD&95735&2&.&543&0&.&004&2&HIPnr,YPC&0&.&52&S&D\\
M004&GJ&65 A&2&.&676&0&.&019&1&YPC&0&.&52&S&D\\
M005&HIP&92403&2&.&965&0&.&017&2&HIPnr,YPC&6&.&07&S&\\
M006&GJ&905&3&.&165&0&.&011&1&YPC&0&.&78&S&D\\
M007&HD&217987&3&.&278&0&.&007&2&YPC,HIPnr&0&.&52&S&D\\
M008&HIP&57548&3&.&354&0&.&015&2&YPC,HIPnr&0&.&53&S&D\\
\end{longtable}

\begin{longtable}{ccr@{ }lr@{ }r@{ }r@{}c@{}l@{\hspace{1em}}r@{}r@{ }r@{ }r@{}c@{}lr@{}c@{}lr@{}c@{}llr@{}c@{}l}
\caption{Component names, positions and proper motions: `Primary' column contains `P' for primary component; `Reference' column gives the reference for position and proper motion; $\rho$ column gives separation of component from primary if larger than $1.0''$. $\rho$ should be considered approximate, and time variable for smaller separations (of order $100\,{\rm AU}$ or less). It is advised to check orbital solutions to find relative positions for a particular epoch.
}\\
\hline
{UNS ID}&{Primary}&\multicolumn{2}{c}{Name}&\multicolumn{11}{c}{Position ICRS 2000.0}&\multicolumn{3}{c}{$\mu_\alpha\cos\delta$}&\multicolumn{3}{c}{$\mu_\delta$}&{References}&\multicolumn{3}{c}{$\rho$}\\
{}&{}&{}&{}&\multicolumn{5}{c}{}&\multicolumn{6}{c}{}&\multicolumn{3}{c}{(mas/yr)}&\multicolumn{3}{c}{(mas/yr)}&{}&\multicolumn{3}{c}{(arcsec)}\\
\hline
\endfirsthead

\caption{Component names, positions and proper motions (continued)}\\
\hline
{UNS ID}&{Primary}&\multicolumn{2}{c}{Name}&\multicolumn{11}{c}{Position ICRS 2000.0}&\multicolumn{3}{c}{$\mu_\alpha\cos\delta$}&\multicolumn{3}{c}{$\mu_\delta$}&{References}&\multicolumn{3}{c}{$\rho$}\\
{}&{}&{}&{}&\multicolumn{5}{c}{}&\multicolumn{6}{c}{}&\multicolumn{3}{c}{(mas/yr)}&\multicolumn{3}{c}{(mas/yr)}&{}&\multicolumn{3}{c}{(arcsec)}\\
\hline
\endhead

\multicolumn{24}{r}{Continued on next page...} \\
\endfoot

\hline
\endlastfoot

% manually reordered F093, F094, A028, A109, A129
% renamed F093: HD 7693 to GJ 55.1 A
% A007 C, A128 C references have second occurence of CCDM removed

M001&P&HIP&87937&17&57&48&.&50&$+$&04&41&35&.&8&-798&.&8&10277&.&3&NLTT&&&\\
M002&P&GJ&406&10&56&28&.&99&$+$&07&00&52&.&0&-3841&.&6&-2725&.&1&LHS&&&\\
M003&P&HD&95735&11&03&20&.&20&$+$&35&58&11&.&6&-577&.&0&-4761&.&8&NLTT&&&\\
M004&P&GJ&65 A&01&39&01&.&54&$-$&17&57&01&.&8&3296&.&2&563&.&9&NLTT&&&\\
M004&&GJ&65 B&01&39&01&.&54&$-$&17&57&01&.&8&3296&.&2&563&.&9&NLTT&&&\\
M005&P&HIP&92403&18&49&49&.&37&$-$&23&50&10&.&4&644&.&2&-192&.&9&NLTT&&&\\
M006&P&GJ&905&23&41&55&.&00&$+$&44&10&38&.&9&100&.&0&-1594&.&1&NLTT&&&\\
M007&P&HD&217987&23&05&52&.&04&$-$&35&51&11&.&1&6767&.&3&1326&.&7&NLTT&&&\\
\end{longtable}

\begin{longtable}{cr@{ }llcr@{}c@{}l@{ $\pm$ }r@{}c@{}lr@{}c@{}l@{ $\pm$ }r@{}c@{}lccc}
\caption{A-K primary spectral types, Tycho photometry and effective temperatures: spectral type and reference; Tycho $B_{\rm T}$,$V_{\rm T}$ magnitudes with standard errors and reference; $T_{\rm eff}$ from gray03 or gray06; $T_{\rm eff}$ computed from Tycho photometry (see text). Where TYC2 and TDSC give the same $B_{\rm T}$,$V_{\rm T}$ and uncertainties we use TYC2 as the reference here. In six cases Tycho photometry is not available, so we give values converted from Johnson $B$,$V$ magnitudes using $V_{\rm T}=V_{\rm J}+{0.090\over0.850}\left(B_{\rm J}-V_{\rm J}\right)$, $B_{\rm T}=V_{\rm J}+{1.090\over0.850}\left(B_{\rm J}-V_{\rm J}\right)$.
}\\
\hline
{UNS ID}&\multicolumn{2}{c}{Primary name}&{SpT}&{Ref.}&\multicolumn{6}{c}{$V_{\rm T}$}&\multicolumn{6}{c}{$B_{\rm T}$}&{Ref.}&{$T_{\rm eff,G}$}&{$T_{\rm eff,T}$}\\
{}&{}&{}&{}&{}&\multicolumn{6}{c}{(mag)}&\multicolumn{6}{c}{(mag)}&{}&{(K)}&{(K)}\\
\hline
\endfirsthead

\caption{A-K primary spectral types, Tycho photometry and effective temperatures (continued)}\\
\hline
{UNS ID}&\multicolumn{2}{c}{Primary name}&{SpT}&{Ref.}&\multicolumn{6}{c}{$V_{\rm T}$}&\multicolumn{6}{c}{$B_{\rm T}$}&{Ref.}&{$T_{\rm eff,G}$}&{$T_{\rm eff,T}$}\\
{}&{}&{}&{}&{}&\multicolumn{6}{c}{(mag)}&\multicolumn{6}{c}{(mag)}&{}&{(K)}&{(K)}\\
\hline
\endhead

\multicolumn{20}{r}{Continued on next page...} \\
\endfoot

\hline
\endlastfoot

% Note: HIC references changed to HIP for simplicity.

K001&HD&22049&K2 V (k)&gray06&3&.&814&0&.&009&4&.&846&0&.&014&TYC2&4999&5005\\
K002&HD&201091&K5 V&gray03&5&.&349&0&.&009&6&.&711&0&.&014&TYC2&&4401\\
K003&HD&209100&K4 V (k)&gray06&4&.&826&0&.&009&6&.&048&0&.&014&TYC2&&4654\\
K004&HD&202560&K7.0&haw96&6&.&845&0&.&011&8&.&476&0&.&017&TYC2&&3915\\
K005&HD&88230&K5&haw95&6&.&751&0&.&010&8&.&340&0&.&016&TYC2&&3990\\
K006&HD&26965&K0.5 V&gray06&4&.&506&0&.&009&5&.&440&0&.&014&TYC2&5124&5199\\
K007&HD&165341&K0- V&gray03&4&.&217&0&.&009&5&.&180&0&.&014&TYC2&5019&5140\\
K008&HD&131977&K4 V&gray06&5&.&880&0&.&010&7&.&163&0&.&016&TYC2&&4544\\
\end{longtable}

\begin{longtable}{cr@{ }llccr@{}c@{}lcr@{}c@{}lc}
\caption{M-type primary spectral types, effective temperatures and Johnson $B,V$ photometry: spectral type and reference, $T_{\rm eff}$ determined from spectral type, Johnson $V$ magnitude and references, Johnson $(B-V)$ colour and references. Where multiple references are given for the photometry, the value given here is the mean of the refereneced values.
}\\
\hline
{UNS ID}&\multicolumn{2}{c}{name}&{SpT}&{SpT ref}&{$T_{\rm eff}({\rm SpT})$}&\multicolumn{3}{c}{$V_{\rm J}$}&{$V_{\rm J}$ ref}&\multicolumn{3}{c}{$(B-V)_{\rm J}$}&{$(B-V)_{\rm J}$ ref}\\
{}&{}&{}&{}&{}&\multicolumn{1}{c}{(K)}&\multicolumn{3}{c}{(mag)}&{}&\multicolumn{3}{c}{(mag)}&{}\\
\hline
\endfirsthead

\caption{M-type primary spectral types, effective temperatures and Johnson $B,V$ photometry (continued)}\\
\hline
{UNS ID}&\multicolumn{2}{c}{name}&{SpT}&{SpT ref}&{$T_{\rm eff}({\rm SpT})$}&\multicolumn{3}{c}{$V_{\rm J}$}&{$V_{\rm J}$ ref}&\multicolumn{3}{c}{$(B-V)_{\rm J}$}&{$(B-V)_{\rm J}$ ref}\\
{}&{}&{}&{}&{}&\multicolumn{1}{c}{(K)}&\multicolumn{3}{c}{(mag)}&{}&\multicolumn{3}{c}{(mag)}&{}\\
\hline
\endhead

\multicolumn{13}{r}{Continued on next page...} \\
\endfoot

\hline
\endlastfoot

% Hack: \hspace{1em} used for first B-V entry to get column roughly centred

M001&HIP&87937&M4&haw95&3100&9&.&55&wei96,bes90&\hspace{1em}1&.&75&wei96,bes90\\
M002&GJ&406&M5.5&haw95&2700&13&.&48&wei96,bes90&2&.&00&wei96,bes90\\
M003&HD&95735&M2&haw95&3400&7&.&50&egg74&1&.&51&egg74\\
M004&GJ&65 A&M5.5&haw95&2700&12&.&57&CNS3&1&.&85&CNS3\\
M005&HIP&92403&M3.5&haw95&3175&10&.&43&wei96,bes90&1&.&75&wei96,bes90\\
M006&GJ&905&M5&haw95&2800&12&.&29&wei96&1&.&91&wei96\\
M007&HD&217987&M0.5&haw96&3700&7&.&37&egg74&1&.&48&egg74\\
M008&HIP&57548&M4&haw95&3100&11&.&14&wei96,bes90&1&.&76&wei96,bes90\\
\end{longtable}

\setlength{\LTcapwidth}{\textheight}

\begin{landscape}

\scriptsize
\begin{longtable}{cr@{ }lrcrrrrrrr@{ }lrrrrc}
\caption{Component cross identifications with common catalogues: system ID, CCDM ID and component, Henry Draper (HD/HDE) ID, Gleise \& Jahreiss (CNS3) ID, Luyten Half Second ID, New Luyten Two Tenths ID (record number in original NLTT), Harvard Revised (BSC5) ID, Positions and Proper Motions ID, Hipparcos ID, Tycho ID, Tycho Double Star Catalogue ID and component, Bonner Durchmusterung ID, Cordoba Durchmusterung ID, Cape Photographic Durchmusterung ID, Yale Parallax Catalogue (PLX) ID, 2MASS Point Source Catalogue ID (these are determined by simple cone search and may not be reliable in some cases).}\\
\hline
{UNS}&\multicolumn{2}{c}{CCDM}&\multicolumn{1}{c}{HD}&{GJ}&\multicolumn{1}{c}{LHS}&\multicolumn{1}{c}{NLTT}&\multicolumn{1}{c}{HR}&\multicolumn{1}{c}{PPM}&\multicolumn{1}{c}{HIP}&\multicolumn{1}{c}{TYC}&\multicolumn{2}{c}{TDSC}&\multicolumn{1}{c}{BD}&\multicolumn{1}{c}{CoD}&\multicolumn{1}{c}{CPD}&\multicolumn{1}{c}{PLX}&\multicolumn{1}{c}{2MASS}\\
%{}&{}&{}&{}&{}&{}&{}&{}&{}&{}&{}&{}&{}&{}&{}&{}&{}&{}\\
\hline
\endfirsthead

\caption{Component cross identifications with common catalogues (continued)}\\
\hline
{UNS}&\multicolumn{2}{c}{CCDM}&\multicolumn{1}{c}{HD}&{GJ}&\multicolumn{1}{c}{LHS}&\multicolumn{1}{c}{NLTT}&\multicolumn{1}{c}{HR}&\multicolumn{1}{c}{PPM}&\multicolumn{1}{c}{HIP}&\multicolumn{1}{c}{TYC}&\multicolumn{2}{c}{TDSC}&\multicolumn{1}{c}{BD}&\multicolumn{1}{c}{CoD}&\multicolumn{1}{c}{CPD}&\multicolumn{1}{c}{PLX}&\multicolumn{1}{c}{2MASS}\\
%{UNS}&{CCDM}&{}&{HD}&{GJ}&{LHS}&{NLTT}&{HR}&{PPM}&{HIP}&{TYC}&\multicolumn{2}{c}{TDSC}&{BD}&{CoD}&{CPD}&{PLX}&{2MASS}\\
%{}&{}&{}&{}&{}&{}&{}&{}&{}&{}&{}&{}&{}&{}&{}&{}&{}&{}\\
\hline
\endhead

\multicolumn{18}{r}{Continued on next page...} \\
\endfoot

\hline
\endlastfoot

M001&&&&699&57&45718&&&87937&425-2502-1&&&+04 3561a&&&4098.00&17574849+0441405\\
M002&&&&406&36&&&&&&&&&&&2553.00&10562886+0700527\\
M003&11033+3558&A&95735&411&37&26105&&75640&54035&2521-2279-1&&&+36 2147&&&2576.00&11032027+3558203\\
M004&&&&65 A&9&5504&&&&&&&&&&343.10&01390120-1757026\\
M004&&&&65 B&10&5505&&&&&&&&&&343.10&01390120-1757026\\
M005&&&&729&3414&47045&&734465&92403&6859-1332-1&&&&-23 14742&&4338.00&18494929-2350101\\
M006&&&&905&549&57692&&&&&&&&&&5736.00&23415498+4410407\\
M007&&&217987&887&70&55782&&400306&114046&7512-1263-1&&&&-36 15693&-36 9694&5584.00&23055131-3551130\\
\end{longtable}

\end{landscape}

\setlength{\LTcapwidth}{\textwidth}

\small
\begin{longtable}{cl}
\caption{Notes for specific systems}\\
\hline
UNS ID & Note \\
\hline
\endfirsthead

\caption{Notes for specific systems (continued)}\\
\hline
UNS ID & Note \\
\hline
\endhead

\multicolumn{2}{r}{Continued on next page...} \\
\endfoot

\hline
\endlastfoot

M009 & Triple system. A,C components are very close binary, B component orbits AC (Delfosse et al. 1999) \\
M011 & CCDM lists a third component (CCDM 00184+4401 C), but this is not associated, \\
     & CCDM 00184+4401 C is TYC 2794-1389-1 \\
M018 & CCDM lists seven other components (CCDM 22281+5741 C,D,E,F,G,H,I), but these are not associated, \\
     & CCDM 22281+5741 C is TYC 3991-30-1, CCDM 22281+5741 D is clearly visible in 2MASS images \\
     & (22 28 10.42 +57 42 44.9), but is not in the PSC, CCDM 22281+5741 E is 2MASS 22281788+5742148, \\
     & CCDM 22281+5741 F is 2MASS 22280456+5742284, identification of CCDM 22281+5741 G,H uncertain, \\
     & CCDM 22281+5741 I is HD 213209 \\
\end{longtable}
\typeout{get arXiv to do 4 passes: Label(s) may have changed. Rerun}
\end{document}